\documentclass[prl,twocolumn,a4paper,groupedaddress,showpacs,floatfix]{revtex4-1}
\usepackage{graphicx,amsmath,amssymb,amsfonts,dsfont}
\newcommand{\mf}[1]{\boldsymbol{#1}}
\newcommand{\ket}[1]{\ensuremath{|#1\rangle}}
\newcommand{\mc}[1]{\ensuremath{\mathcal{#1}}}
\newcommand{\bra}[1]{\ensuremath{\langle #1 |}}

\newcommand{\ve}{\varepsilon}

\newcommand{\mean}[1]{\ensuremath{ \langle #1  \rangle}}

\usepackage{color}

\begin{document}

\title{Magnetic monopoles and synthetic spin-orbit coupling in Rydberg macrodimers}

\author{Martin Kiffner${}^{1,2}$}
\author{Wenhui Li${}^{1,3}$}
\author{Dieter Jaksch${}^{2,1}$}

\affiliation{Centre for Quantum Technologies, National University of Singapore, 3 Science Drive 2, Singapore 117543${}^1$}
\affiliation{Clarendon Laboratory, University of Oxford, Parks Road, Oxford OX1 3PU, United Kingdom${}^2$}
\affiliation{Department of Physics, National University of Singapore, 117542, Singapore${}^3$}









\begin{abstract}
We show that sizeable Abelian and non-Abelian gauge fields arise in 
the relative motion of two dipole-dipole interacting Rydberg atoms. Our system exhibits two magnetic monopoles for 
adiabatic motion in one internal  two-atom state. These monopoles     
occur at a characteristic distance between the atoms that is of the order of one micron. 
The deflection of the relative motion due to the Lorentz force gives rise to a clear signature of 
the effective magnetic field. 
In addition, we consider non-adiabatic transitions between two near-degenerate internal states and 
show that the  associated gauge fields are non-Abelian. 
We present  quantum mechanical calculations of this synthetic spin-orbit coupling and 
show that it realizes a velocity-dependent beamsplitter. 
\end{abstract}

\maketitle

%
%
Gauge theories represent a cornerstone of modern physics and play a prominent role in
classical and quantum electrodynamics, the standard model of elementary particle physics and
condensed matter physics. In view of the importance of this concept tremendous effort
has been made to create artificial gauge fields for neutral 
atoms~\cite{ruseckas:05,dalibard:11,dum:96,lin:09,lin:09n,lin:11,aidelsburger:11,jaksch:03,struck:11,struck:12,jimenez_garcia:12,hauke:12}
and to investigate the resulting atom dynamics in the quantum regime. 
In these schemes, engineered light-matter interactions cause neutral atoms to behave like charged particles in an electromagnetic field.
 
Artificial gauge fields allow the simulation of theoretical models that are otherwise inaccessible. 
For example, the realization of magnetic monopoles affecting the relative nuclear motion of diatomic molecules 
has been discussed in~\cite{moody:86}. However, gauge field effects in molecules are 
usually very small since they arise from terms that are neglected 
in the Born-Oppenheimer approximation (BOA) \cite{born:27} which is very well satisfied in many molecular systems. 
In addition, the experimental observation of these effects is considerably hampered by the small size of conventional molecules.
 
Recently, extremely large molecules comprised of  two Rydberg atoms with non-overlapping electron clouds
have been proposed~\cite{boisseau:02,samboy:11,kiffner:12} and observed~\cite{overstreet:09}.
Typical internuclear spacings exceed $1\,\mu\text{m}$, and thus  the experimental observation of
Rydberg-Rydberg correlations becomes feasible~\cite{schwarzkopf:11,schauss:12,gaetan:09,urban:09}. 
These so-called macrodimers  interact via well-understood and controllable dipole-dipole potentials. 
Importantly, the validity of the BOA cannot be established via the mass ratio of nuclei and electrons for these systems.

\begin{figure}[t!]
\includegraphics[width=8.5cm]{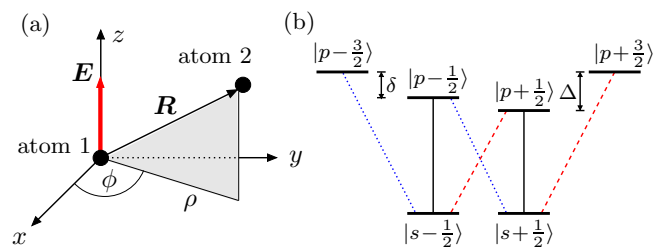}
\caption{\label{fig1}
(Color online) (a) System geometry of two dipole-dipole interacting Rydberg atoms.  $\mf{R}$ is the relative position  of atom 2 with respect to atom 1, 
and $\rho$ is the distance of atom 2 from the $z$ axis. 
The $z$ direction is distinguished by a DC electric field $\mf{E}$. 
(b) Internal level structure of each Rydberg atom.   The Stark shifts $\delta$ 
and $\Delta$ are negative. We assume $\delta\not=\Delta$. 
The dipole transitions  indicated by blue dotted, black solid 
and red dashed lines couple to  $\sigma^-$, $\pi$ and $\sigma^+$ polarized fields, respectively.
}
\end{figure}

Here we show that  dipole-dipole interacting Rydberg atoms can exhibit Abelian and non-Abelian gauge fields that 
influence the quantum dynamics of the relative atomic motion substantially. 
In contrast to the Rydberg macrodimer proposal in~\cite{kiffner:12}, the system  in Fig.~\ref{fig1} 
is distinguished by  an asymmetric Stark shift of the  Zeeman sublevels. 
We find that this broken symmetry gives rise to  magnetic monopoles if 
the system evolves adiabatically in one internal two-atom state. 
These monopoles occur at a characteristic distance between the atoms that is of the order of one micron. 
The Lorentz force associated with the magnetic field near a monopole 
results in a sizeable deflection of the  relative atomic motion. 
This effect can be interpreted in terms of an exchange between orbital and internal spin angular momentum 
as the internal molecular state changes while the atoms move. 
Moreover, we investigate  non-adiabatic transitions between two internal two-atom states and find that 
the associated gauge fields are are non-Abelian.  
This synthetic spin-orbit coupling  creates a coherent superposition of two spatial configurations of the atoms. 
We expect that our findings are relevant for other dipole-dipole interacting 
systems like  polar molecules and magnetic atoms.
%
%

%
The geometry of the two atom system under consideration is shown in Fig.~\ref{fig1}(a). 
In order to account for the azimuthal symmetry of the system, we express the relative position  $\mf{R}$  
of atom 2 with respect to atom 1 in terms of cylindrical coordinates, 
$\mf{R} = (\rho \cos\phi,\rho \sin\phi, z)$. 
We omit the center-of-mass motion which is uniform and investigate the 
relative motion of the two dipole-dipole interacting Rydberg atoms. 
The  Hamiltonian of this system is given by 
\begin{align}
H = H_{\text{R}} + H_{\text{S}} + V_{\text{dd}},
\label{H}
\end{align}
where 
$H_{\text{R}} = \mf{p}^2/(2\mu)$ 
is the kinetic energy of the relative motion and $\mu$ is the reduced mass. 
$H_{\text{S}}$ describes the internal levels of the two uncoupled atoms and $V_{\text{dd}}$ is the dipole-dipole interaction~
\cite{kiffner:12}.
In each Rydberg atom we consider two angular momentum multiplets as shown in Fig.~\ref{fig1}(b). 
The lower $ns_{1/2}$ states have total angular momentum $J=1/2$, 
and the excited multiplet is comprised of  $np_{3/2}$ states with total angular momentum $J=3/2$. We specify the individual 
atomic states  $\ket{\ell m_j}$ by their orbital angular momentum $\ell$ and azimuthal total angular momentum $m_j$. 
A DC  electric field $\mf{E}$ in the $z$ direction  defines the quantization axis and gives rise to 
Stark shifts of the magnetic sublevels. We assume that the Stark shifts are different  in the $m_j>0$ 
and $m_j<0$ manifolds, which could be achieved, e.g., by inducing additional AC stark shifts. 
For simplicity  we focus on the level scheme shown in  Fig.~\ref{fig1}(b), where the asymmetry is characterized 
by the ratio $\Delta/\delta$ of the Stark shifts $\delta$ and $\Delta$. 
The relevant subspace of two-atom states is spanned by the $N=16$ $nsnp$ states where one atom is in a $ns_{1/2}$ state and the other 
in a $np_{3/2}$ state. 
For every value of $\mf{R}$ we introduce a set of orthonormal 
eigenstates of the Hamiltonian $H_{\text{S}} + V_{\text{dd}}$, 
\begin{align}
(H_{\text{S}} + V_{\text{dd}}) \ket{\psi_i(\mf{R})} = \epsilon_i (\mf{R}) \ket{\psi_i(\mf{R})}, 
\label{eigen}
\end{align}
where $\epsilon_i (\mf{R})$ are the corresponding eigenvalues. 
With these definitions, the   full quantum state of the two-atom system can be  written as 
$\ket{\Psi} = \sum_{i=1}^{N}\int \text{d}^3R \; \alpha_i(\mf{R}) \ket{\psi_i(\mf{R})}\otimes\ket{\mf{R}}$. 
Next we assume that the dynamics is confined to  
$q$ eigenstates of $H_{\text{S}} + V_{\text{dd}}$, i.e., there may be non-adiabatic transitions within 
the first $q$ eigenstates, but transitions to other states $\ket{\psi_l(\mf{R})}$ ($l>q$) can be neglected. 
We follow the procedure described in~\cite{wilczek:84,dum:96,ruseckas:05,dalibard:11} and derive from Eq.~(\ref{H}) 
an effective Schr\"odinger equation for the  wavefunctions 
$\mf{\mc{\alpha}} = (\alpha_1,\ldots,\alpha_q)$, $q < N$,
\begin{align}
i\hbar \partial_t \mf{\alpha} = \left[\frac{1}{2\mu}(\mf{p} -\mf{A})^2 + V + \Phi \right] \mf{\alpha}. 
\label{s2}
\end{align}
Equation~(\ref{s2}) is equivalent to the Schr\"odinger equation  of a charged particle 
in an electromagnetic field, characterized by the vector potential $\mf{A}$ and scalar potential $\Phi$.  
Here $V$ and $\mf{A}$ are $q\times q$ matrices whose matrix elements  for $k,l\le q$ are 
given by 
\begin{align}
& V_{kl}  = \delta_{kl} \epsilon_k (\mf{R}) , \quad 
\mf{A}_{kl}   = i \hbar \bra{\psi_k(\mf{R})}\nabla\ket{\psi_l(\mf{R})} , \label{adef}
\end{align}
and $\delta_{kl}$ is the Kronecker delta. 
We find that the impact of the scalar potential $\Phi$  on the presented results is negligible, 
and thus  omit it in the following.  
The Cartesian components $B^{(i)}$ ($i\in \{1,2,3\}$) of the artificial magnetic field are defined as
\begin{align}
B^{(i)}& = \frac{1}{2} \ve_{ikl} F^{(kl)}, \label{bfeld} \\
F^{(kl)}& = \partial_k A^{(l)} -\partial_l A^{(k)}  -\frac{\imath}{\hbar}\left[A^{(k)},A^{(l)} \right], \label{Ftensor} 
\end{align}
where $\ve_{ikl}$ is the Levi-Civita tensor and we employed Einstein's sum convention. 
The $q\times q$ matrices $A^{(i)}$  describe  non-Abelian gauge fields if the commutator 
$[A^{(k)},A^{(l)} ]$ is different from zero.   
The magnetic field gives rise to a Lorentz force  which is proportional to the velocity
$\mf{v}=(\mf{p}-\mf{A})/\mu$ of the relative motion. 
%
\begin{figure}[t!]
\includegraphics[width=8.6cm]{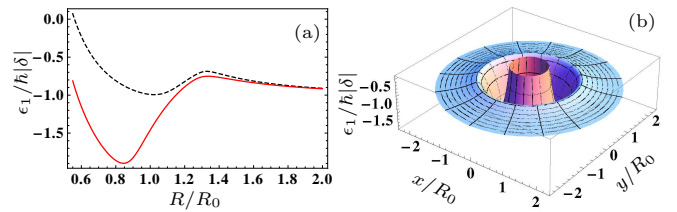}
\caption{\label{fig2}
(Color online) (a) Potential curve $\epsilon_1$ in the $x-y$ plane as a 
function of scaled internuclear spacing $R/R_0$. 
The parameters  are $\Delta=-1.5|\delta|$ (dashed black line) and $\Delta=-3|\delta|$ (red solid line).
(b)  Potential curve $\epsilon_1$   in the $x-y$ plane for $\Delta=-3|\delta|$. 
}
\end{figure}
%

\begin{figure}[t!]
\includegraphics[width=8.6cm]{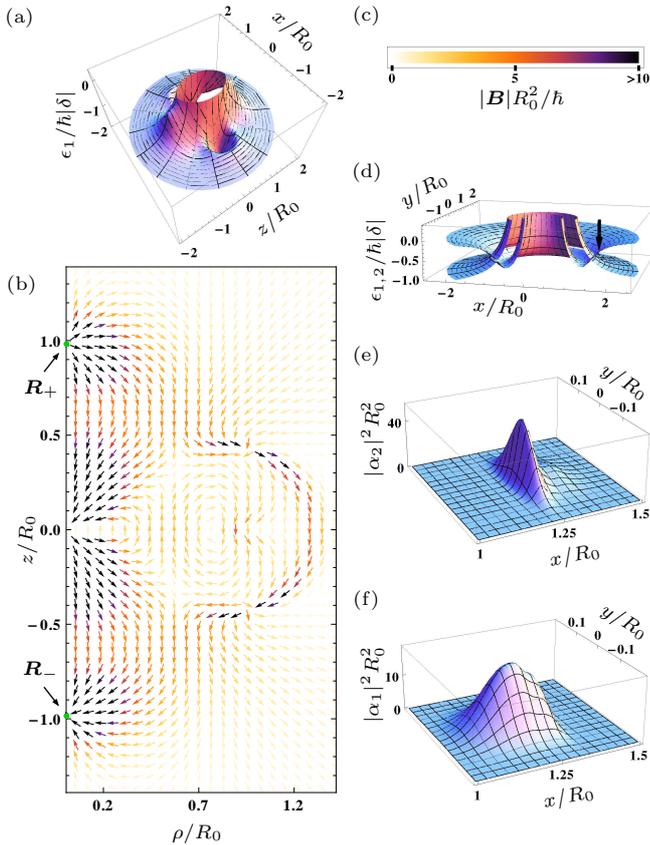}
\caption{\label{fig3}
(Color online) 
(a)-(c) Abelian case. 
We consider adiabatic motion in  state $\ket{\psi_1}$. 
(a) Potential curve $\epsilon_1$   in the $x-z$ plane for $\Delta=-3|\delta|$. 
(b) Artificial magnetic field $\mf{B}$ in the $\rho-z$ plane. $\mf{R}_{\pm}$ indicate 
the positions of the magnetic monopoles. 
(c) Color code for the absolute value of the effective magnetic field in (b). 
(d)-(f) Non-Abelian case. 
We consider non-adiabatic motion in states $\ket{\psi_1}$ and $\ket{\psi_2}$ for $\Delta=-1.16|\delta|$. 
(d) Potential curves $\epsilon_1$  and $\epsilon_2$ in the $x-y$ plane. 
The initial position of the Gaussian wavepacket in the upper well $\epsilon_2$ is indicated by a black arrow. 
(e) and (f) show the probability density of the evolved  wave functions $\alpha_1$  and $\alpha_2$ at $t|\delta| = 220$. 
The population in the upper (lower) well state $\ket{\psi_2}$ ($\ket{\psi_1}$) is $52.7\,\%$ ($47.3\,\%$).
In (e) and (f), we  consider ${}^{39}$K atoms with principal quantum number $n=20$, $|\delta|= 2\pi \times 24.7\,\text{MHz}$  
and $R_0=1.28 \mu\text{m}$. 
}
\end{figure}

%
The eigenstates $\ket{\psi_i}$ and eigenvalues $\epsilon_i$  of the 
Hamiltonian $H_{\text{S}} + V_{\text{dd}}$ in Eq.~(\ref{eigen}) 
can be obtained numerically. Here we focus  on one particular potential curve 
that exhibits a  potential well such that the two dipole-dipole interacting Rydberg atoms can form 
a giant molecule. This  potential curve is labelled by $\epsilon_1$ and shown in Fig.~\ref{fig2}(a) 
for different ratios of $\Delta/\delta$.    
The potential minimum occurs roughly at  the characteristic length 
$R_0=(|\mathcal{D}|^2/(4\pi \epsilon_0\hbar |\delta|)^{1/3}$  
denoting the distance where the magnitude of the dipole-dipole interaction equals the Stark splitting $\hbar|\delta|$ 
~\cite{kiffner:12}, and $|\mathcal{D}|$ is the reduced dipole matrix element of the $ns\leftrightarrow np$ transition~\cite{kiffner:12}. 
Due to the azimuthal symmetry of the system a donut shaped potential well arises in the $x-y$ plane [see Fig.~\ref{fig2}(b)]. 
The dependence of $\epsilon_1$ on $x$ and $z$ is displayed in Fig.~\ref{fig3}(a), 
showing that the width of the potential well in $z$ direction 
is comparable to its width in the $x-y$ plane. 
Note that a potential well with similar features but for $\delta=\Delta < 0$ was reported in~\cite{kiffner:12}. 
Next we consider the case $q=1$ in Eq.~(\ref{s2}) and 
consider the adiabatic motion in the  eigenstate $\ket{\psi_1}$ corresponding to $\epsilon_1$. 
We choose the phase of $\ket{\psi_1}$ such that the vector potential  $\mf{A}_1=\mf{A}_{11}$ 
obeys the Coulomb gauge ($\text{div}\mf{A}_1 = 0$)~\cite{jpb}. It follows that 
\begin{align}
A_1^{(\phi)}(\rho,z)=\mf{A}_1\cdot\mf{e}_{\phi} = \frac{1}{{\rho}}\bra{\psi_1(\mf{R})} J_z \ket{\psi_1(\mf{R})} \label{insightS2}
\end{align}
is the only non-zero component of the vector potential. In this equation, 
$J_z = J_z^{(1)} + J_z^{(2)}$ and $J_z^{(\alpha)}$ is the $z$ component of the total angular momentum operator 
of the internal states of atom $\alpha$ and $\mf{e}_{\phi}$ is the unit vector in $\phi$ direction. 
%
%
The spatial variation of the vector potential determines the magnetic field $\mf{B}$ according to  Eq.~(\ref{bfeld}). 
We find that $\mf{B}$ is only different from zero if $\delta\not=\Delta$  such that the 
symmetry of the system is broken. In contrast, 
for a symmetric level scheme the expectation value of $J_z$ in Eq.~(\ref{insightS2}) yields zero. 
Figure~\ref{fig3}(b) shows the magnetic field in the $\rho-z$ plane for $\Delta=-3|\delta|$. 
The most remarkable features of $\mf{B}$ are a source and a drain of magnetic flux   near $\mf{R}_+\approx + 0.98 R_0\mf{e}_z$ and 
$\mf{R}_-\approx - 0.98 R_0\mf{e}_z$, respectively. 
We integrate the magnetic field over a sphere $S_{\pm}$ centered at $\mf{R}_{\pm}$ and evaluate  the Chern number~\cite{xiao:10} 
$\mc{C}_{\pm} = \int_{S_{\pm}}\mf{B}\cdot \text{d}\mf{S}/(2\pi)$.  The result is  $\mc{C}_{\pm}=\pm 1$, demonstrating 
that our system exhibits two magnetic monopoles on the $z$ axis. 
We emphasize that the  monopoles arise at an atomic separation roughly given by $R_0$. This parameter can be controlled 
by the magnitude of the dipole-dipole interaction and the Stark splitting, and  is typically of the order of one micron. 

Next we show that  the artificial magnetic field gives rise to a sizeable deflection of the relative atomic motion. 
To this end, we suppose that 
the relative position of the two atoms is  initially given by $\mf{R}_{\text{I}}=R_0(-1.5 \mf{e}_z+0.05\mf{e}_x)$. 
The initial velocity of the relative motion is $\mean{\mf{v}}=v_0 \mf{e}_z$ with $v_0>0$ such that 
the atoms move towards the region of strong magnetic fields near $\mf{R}_-$.  
Here we treat the relative atomic motion classically, which is justified if the wavepacket associated with 
the relative motion is very well localized. Since $\mf{R}_{\text{I}}$ contains a small offset in the positive $x$ direction, 
the potential curve $\epsilon_1$ will result in a deflection in the positive $x$ direction [see Fig.~\ref{fig3}(a)]. 
Note that our choice of  initial conditions and the azimuthal symmetry of the system imply that 
the motion remains in the $x-z$ plane if there  were no magnetic fields. 
However, the magnetic field pointing towards $\mf{R}_-$ yields to a deflection $y_-$ in the negative $y$ direction 
via the Lorentz force. 
On the contrary, the magnetic field will have the opposite effect if we mirror our initial conditions 
at the $y-z$ plane, i.e., for $\mf{\tilde{R}}_{\text{I}}=R_0(-1.5 \mf{e}_z - 0.05\mf{e}_x)$.  
In this case, the relative motion remains in the $x<0$ half plane and the Lorentz force results in a deflection 
$y_+$ in the positive $y$ direction. 
It follows that the difference $\Delta y=y_+-y_-$  is a direct measure of the effective magnetic field. In 
addition, it reflects the  broken chiral symmetry arising from the asymmetric level scheme in Fig.~\ref{fig1}(b). 

In order to obtain a quantitative description of this effect, we derive a set of coupled equations for the mean values 
$\mean{\mf{R}}$ and $\mean{\mf{v}}$ from the Hamiltonian $H$ in Eq.~(\ref{H})~\cite{tannoudji:lc}. 
We  consider ${}^{23}$Na atoms with principal quantum number 
$n=15$ and $|\delta|= 2\pi \times 39.0\,\text{MHz}$~\cite{comment1}. 
This yields $R_0=0.75 \mu\text{m}$, and hence the distance between the atoms is initially given by 
$|\mf{R}_{\text{I}}|=|\mf{\tilde{R}}_{\text{I}}|\approx 1.13\mu\text{m}$. Furthermore, we set $v_0=195\hbar/(R_0\mu)$. 
We neglect effects due to the finite lifetime of the molecule ($T|\delta|\approx540$) and thus restrict 
the analysis to times $t\le T$. 
From  semi-classical simulations for the two initial positions  $\mf{R}_{\text{I}}$ and 
$\mf{\tilde{R}}_{\text{I}}$ we find   $\Delta y = 0.1 R_0$ for $t|\delta|=238$, 
and  $\Delta y = 0.3 R_0$ at  $t|\delta|=483$. It follows that the magnetic 
field results in a substantial deflection of the relative atomic motion. 
Note that our simulations allow us to confirm that the motion remains adiabatic at all times.  

\begin{figure}[t!]
\includegraphics[width=8.6cm]{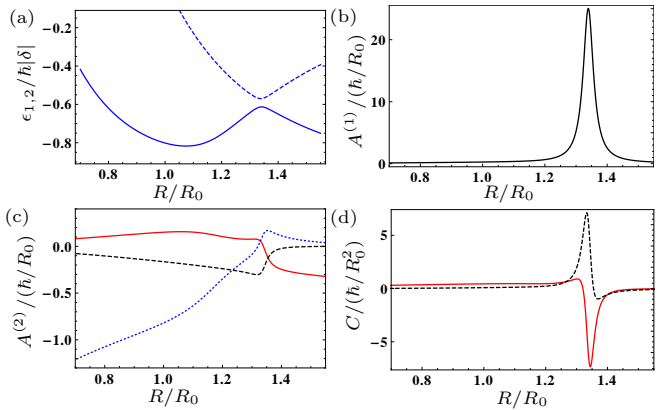}
\caption{\label{fig4}
(Color online) 
(a) Potential curves $\epsilon_1$ (solid blue line) and $\epsilon_2$ (dashed blue line) in the $x-y$ plane. 
(b) Imaginary part of  $A_{12}^{(1)}=[A_{21}^{(1)}]^*$ for $\phi=0$. 
(c) Real parts  of $A_{11}^{(2)}$ (solid red line), $A_{22}^{(2)}$ (dashed black line) and 
$A_{12}^{(2)}=[A_{21}^{(2)}]^*$ (dotted blue line) for $\phi=0$. 
(d) Matrix elements  of the commutator $C=\imath  [A^{(1)},A^{(2)}]/\hbar$. The red solid line shows 
$C_{11}=-C_{22}$, and the black dashed line represents $C_{12}=C_{21}$. 
In (a)-(d), we set $\Delta=-1.16|\delta|$. All components of $\mf{A}$ that are not shown in (b) and (c) are zero. 
}
\end{figure}

Next we show that the vector potential $\mf{A}$ can give rise to 
a coupling between the relative atomic motion and internal electronic states. 
In order to demonstrate this synthetic spin-orbit coupling, we consider an additional 
potential curve $\epsilon_2$ with  corresponding state $\ket{\psi_2}$. 
Here we focus on the two-dimensional setting where the motion is confined to the $x-y$ plane. 
The two potential curves 
$\epsilon_1$ and $\epsilon_2$ become near-degenerate for $R\approx 1.33 R_0$ and $\Delta=-1.16|\delta|$ and are shown in 
Fig.~\ref{fig4}(a). In order to investigate the quantum dynamics in the two cylindrically symmetric 
potential wells,  we evaluate the vector potential  $\mf{A}$ in Eq.~(\ref{adef}) by numerical means~\cite{jpb}. 
Note that $\mf{A}$ is now represented by a $2\times2$ matrix, where each component 
$\mf{A}_{kl}$ is a 3-column vector. We find that the  component $A^{(3)}$ is zero, and the 
non-zero parts  of $A^{(1)}$ and $A^{(2)}$ are shown in Figs.~\ref{fig4}(b) and~(c), respectively. 
All components of $\mf{A}$ are evaluated for $\phi=0$ such that 
$A^{(1)}$ ($A^{(2)}$) can be identified with the radial (azimuthal) component of $\mf{A}$. 
%
%
Near the avoided crossing the  off-diagonal element $\mf{A}_{12}$ 
can induce non-adiabatic transitions between the states $\ket{\psi_1}$ and $\ket{\psi_2}$. 
The coupling strength depends on the energy difference $|\epsilon_1-\epsilon_2|$ and 
the velocity of the relative motion.  
For a quantitative description of this synthetic spin-orbit coupling, we assume that the system is initially at 
rest and prepared in the upper well state $\ket{\psi_2}$ [see Fig.~\ref{fig3}(d)]. 
We model the wavepacket corresponding to the relative atomic motion by a Gaussian 
with a full width at half maximum of $\sigma\approx 75\,\text{nm}$ centered at $R=1.5 R_0$ 
and solve Eq.~(\ref{s2}) for $q=2$ in a box with radius $2.2R_0$. 
As the system evolves, it will oscillate in the upper well, and  near the avoided crossing some population will be coherently 
transferred to the lower well state $\ket{\psi_1}$. The probability densities in the two states after the avoided 
crossing has been traversed once is shown in Figs.~\ref{fig3}(e) and~(f). For the chosen parameters an almost 
equal superposition of the two internal states is created. Note that the two wavepackets experience 
different potentials [see Fig.~\ref{fig4}(a)] and hence they will separate in space for longer evolution times. 
We emphasize that the gauge fields $A^{(1)}$ and $A^{(2)}$ are  strongly non-Abelian 
as shown in  Fig.~\ref{fig4}(d). 
The commutator $C=\imath  [A^{(1)},A^{(2)}]/\hbar$ is of  the same order of magnitude as 
the first term in Eq.~(\ref{Ftensor}), and thus the non-Abelian signature  is 
significant whenever the magnetic field gives rise to sizeable effects in the quantum dynamics of the system. 
This opens up the possibility to study the rich physics resulting from non-Abelian gauge fields~\cite{jacob:07}, 
which  is subject to further investigation. 

In summary, we have shown that the dipole-dipole interaction between Rydberg atoms 
can induce Abelian and non-Abelian artificial gauge fields that influence the 
relative atomic motion significantly. 
The experimental realization of  our scheme could be achieved in optical lattices 
where the lattice constant matches the desired initial separation of the atoms. 
Alternatively, one could start with 
a similar setup as described in~\cite{gaetan:09,urban:09}, where the dipole-dipole interaction 
between two individual Rydberg atoms trapped in optical tweezers was investigated. 
The optical potentials allow one to control the initial position of the atoms before they are excited 
to the diatomic $nsns$ state via laser fields. A subsequent microwave field prepares the system in 
the desired $nsnp$ state $\ket{\psi_1}$ or $\ket{\psi_2}$. In addition, the optical trapping potentials could 
transfer linear momentum  to the atoms before the excitation to the Rydberg states occurs. 
Our calculations for the deflection in the monopole field were carried out at zero temperature. 
By considering a thermal velocity distribution, we estimate that the deflection pattern 
will be washed out if the temperature exceeds approximately $100\,\text{nK}$. 
These temperatures are routinely achieved in optical lattices and dipole traps~\cite{arnold:11}. 
Finally, the observation of the relative atomic motion requires measurements  of the density-density correlations 
of the two Rydberg atoms. Such measurements have been performed by ionization of the Rydberg atoms~\cite{schwarzkopf:11} 
and by de-excitation to the ground state followed by advanced imaging techniques~\cite{schauss:12}. 
We thus believe that the experimental observation of the deflection in the monopole field and  
the splitting of the motional wavepacket is feasible with current or next-generation imaging techniques. 


\begin{thebibliography}{10}
\bibitem{comment1} The moment of inertia $I=\mu R_0^2$ increases with principal quantum number $n$. 
For our setup we find that values of $n\approx 20$ maximize the mechanical effects of the artificial 
gauge fields on the relative atomic motion.
\end{thebibliography}
\end{document}